\RequirePackage{lineno}
\linenumbers

\documentclass[12pt]{iopart}
\usepackage{iopams}  
\usepackage{multirow}
\usepackage{graphicx}
\usepackage{epsfig}
\begin{document}

\title{Highlights from PHENIX - II}

\author{Terry C. Awes  (for the PHENIX Collaboration
\footnote{A list of members of the PHENIX Collaboration can be found at the end of this issue})}
\address{Oak Ridge National Laboratory, Oak Ridge, TN 37831 USA}
\ead{awes@mail.phy.ornl.gov}
\begin{abstract}

This contribution highlights recent results from the PHENIX Collaboration at RHIC with emphasis on those obtained through lepton and photon measurements in PHENIX.

\end{abstract}

\submitto{\JPG}

\section{Introduction}

Following the discovery of the surprising "perfect liquid" properties of the dense matter being produced in heavy ion collisions at RHIC~\cite{perfectliquid},  effort has intensified to quantify the detailed characteristics of that matter.
Photons and leptons are of  special  interest because they are penetrating probes that do not  undergo strong interactions and therefore are unlikely to interact with the dense matter after their production. Thus, they carry information about the system at the time of their production, throughout the entire evolution of the collision. 
The PHENIX experiment at RHIC was designed with emphasis on the measurement of  leptons and photons, with electron and photon identification at mid-rapidity, and muon spectrometers at forward and backward rapidities (see~\cite{Franz} for a description of PHENIX). 

As a result of their weaker electromagnetic coupling, lepton and photon production are rare processes that require large data samples for precise measurements.  As a consequence of their low production rate the measurements are also subject to large backgrounds. In the case of directly radiated  photons, the backgrounds are mostly photons from radiative decays of  long-lived neutral mesons, predominantly the abundantly produced neutral pions. In the case of  the electron measurement the background is mostly the internal or external conversion of these radiative decay photons into electron-position pairs, and in the case of the muon measurements it is the weak decay of charged pions into muons. 

With the $8^{th}$ RHIC Run period recently completed, many of the new results reported by PHENIX are improved measurements of  previously published results obtained from the smaller data sets of the earlier RHIC runs. This is particularly true of the lepton and photon measurements.  As an example, due to the factor of 10 increase in the PHENIX Run 4 data sample compared to Run 2, the measurement of the neutral pion spectra in Au+Au collisions has been extended to nearly 20 GeV/c transverse momentum with decreased statistical and systematic errors~\cite{ppg080}.  This directly translates into improved direct photon and non-photonic electron measurements. 

\section{Parton Energy Loss} \label{sec:eloss}

One of the most exciting early results from RHIC was the observed strong suppression of  neutral pion production in central Au+Au collisions~\cite{PHsup} compared to expectations from scaled p+p collisions, and the lack of suppression of the direct photon yield~\cite{PHgam}. This strongly supported the conclusion that the initial collisions occurred at the expected rate, as evidenced by the expected direct photon yield, but that the neutral pions were suppressed due to strong interactions and energy loss of the initially scattered parton as it traversed the dense medium prior to fragmentation into particles like the pion.

\begin{figure}[h]
	\begin{minipage}[t]{0.5\linewidth}
		\centering
		\includegraphics[width=\linewidth,height=7.0cm]{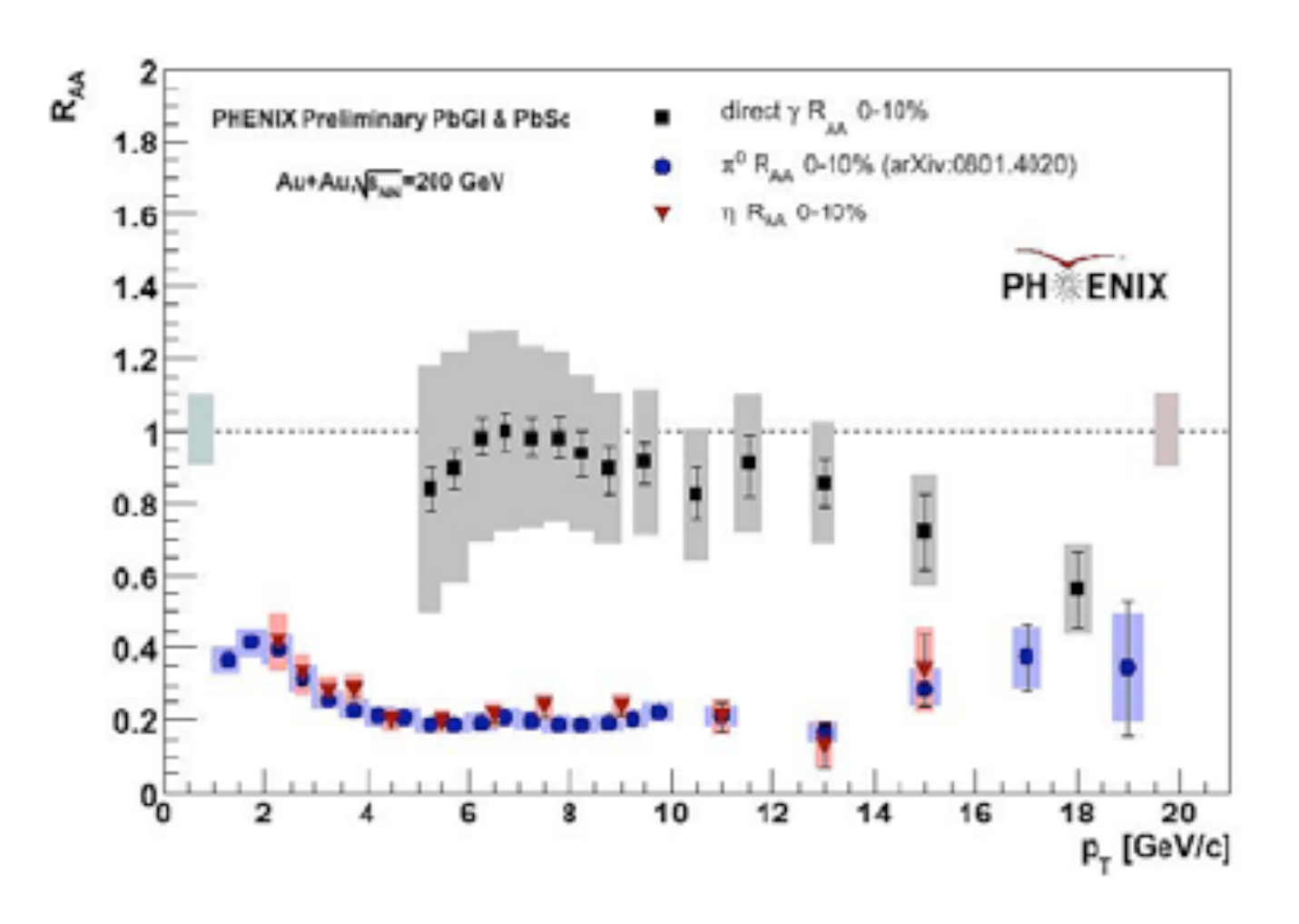}
		\caption{{Nuclear modification factor R$_{AA}$ for $\pi^0, \eta,$ and direct $\gamma$ production as a function of transverse momentum for central Au+Au collisions.}}
		\label{fig:AuRAA}
	\end{minipage} 
	\begin{minipage}[t]{0.5\linewidth}
		\centering
		\includegraphics[width=\linewidth,height=7.0cm]{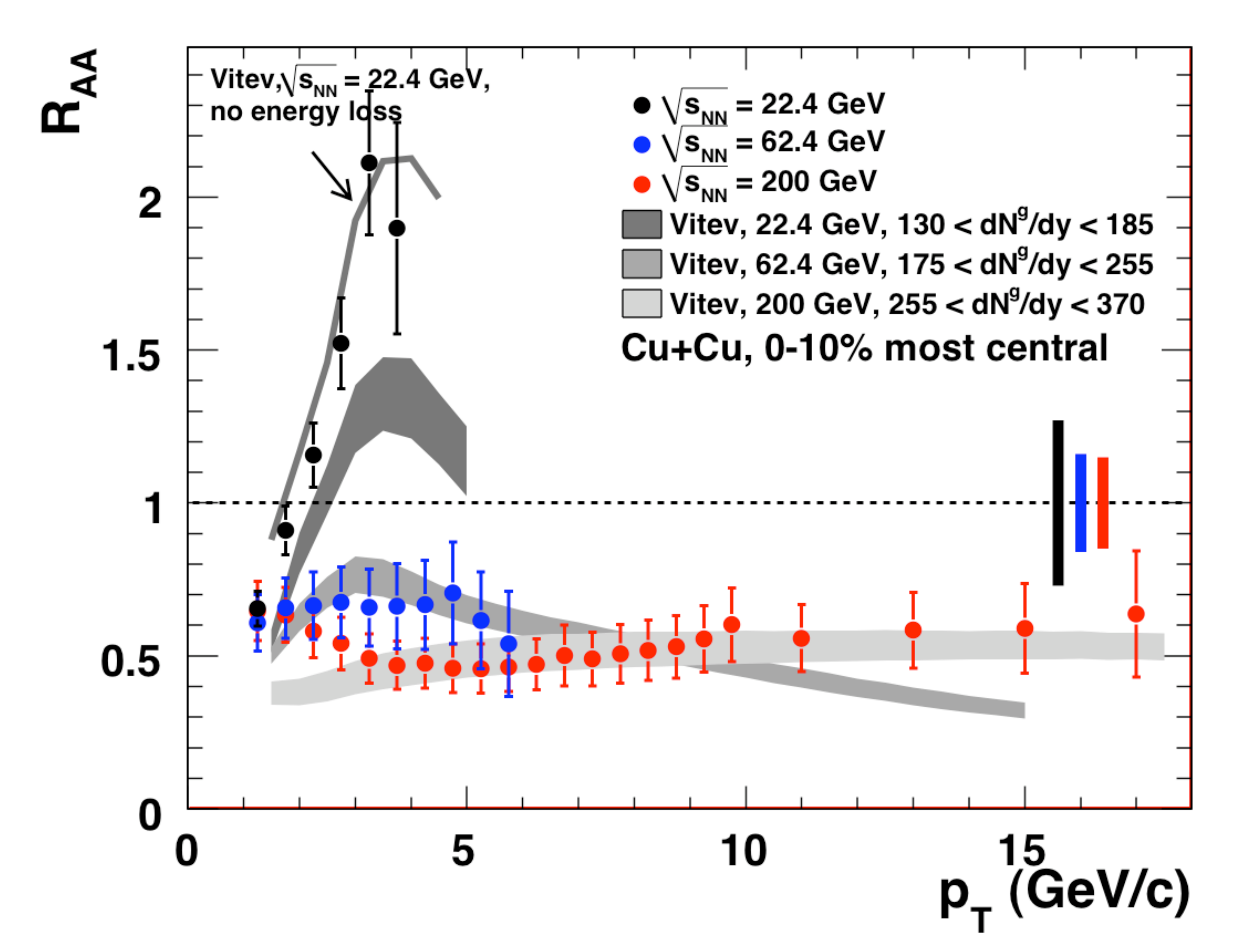}
		\caption{{Nuclear modification factor R$_{AA}$ for $\pi^0$ production in Cu+Cu collisions 
		at $\sqrt{s_{NN}}=22.4,$ 62.4, and 200 GeV. The shaded regions are parton energy loss 
		model predictions~\cite{Vitev}. }} 
		\label{fig:CuRAA}
	\end{minipage}
\end{figure}

Nuclear effects are quantified in terms of the nuclear modification factor $R_{AB}$ defined as 
$R_{AB} = (dN/dp_T |_{A+B}) /  (\langle N_{coll} \rangle \times dN/dp_T |_{p+p})$ where  $\langle N_{coll} \rangle$ is  the average number of independent nucleon-nucleon collisions for the selected class of events.
Final results from analysis of the Run 4 data set shown in Figure~\ref{fig:AuRAA} demonstrate that the $\pi^0$ suppression, and $\eta$ suppression as well, is consistent with being constant with transverse momentum over the region from  5 to 20 GeV/c~\cite{ppg080}. 

\begin{figure}[h]
	\begin{minipage}[t]{0.5\linewidth}
		\centering
		\includegraphics[width=\linewidth,height=7.0cm]{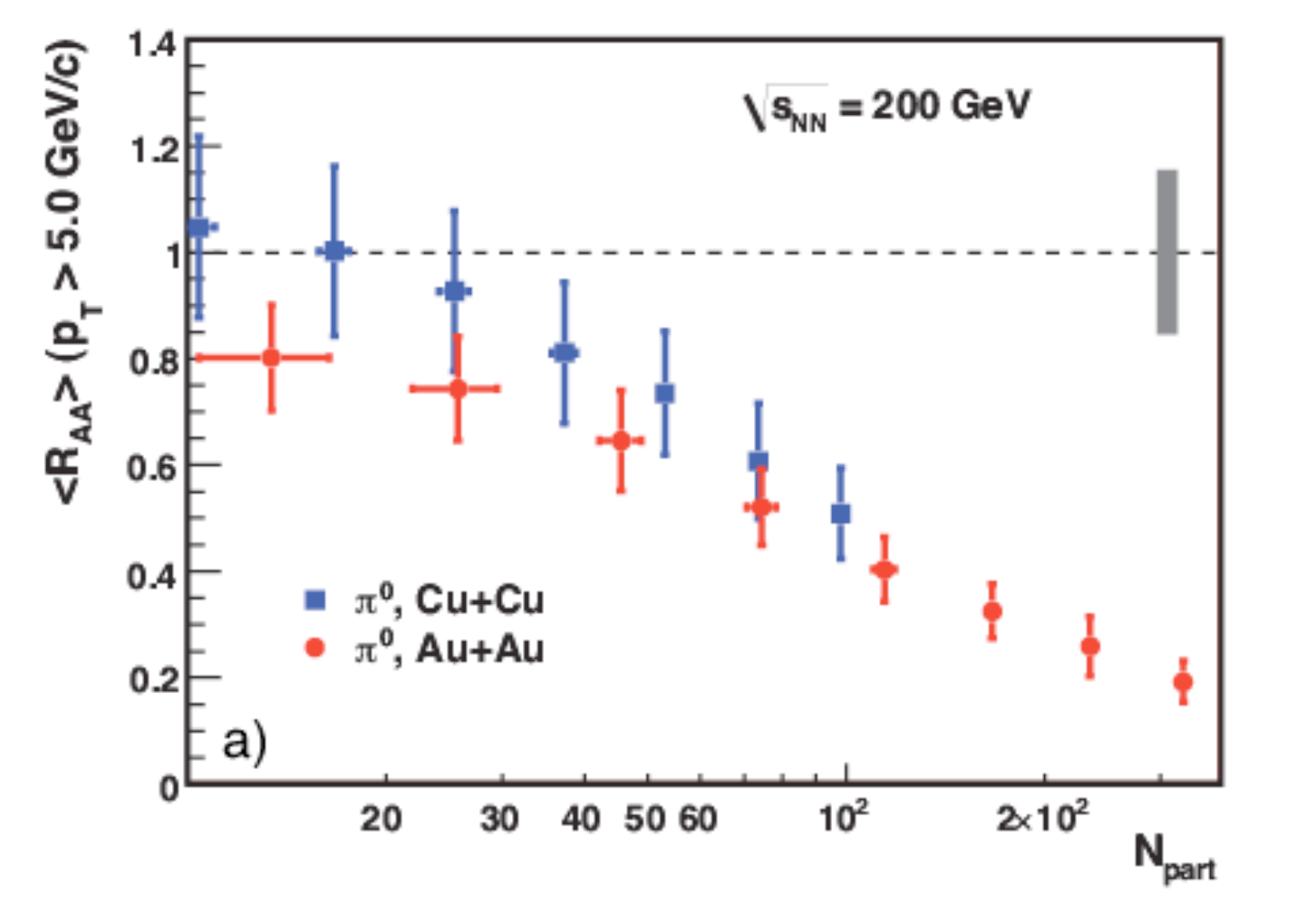}
		\caption{{Nuclear modification factor R$_{AA}$ for high $p_T \pi^0$ production as a function of the number of participant nucleons in Cu+Cu and Au+Au collisions.}}
		\label{fig:CuAuRAA}
	\end{minipage} 
	\begin{minipage}[t]{0.5\linewidth}
		\centering
		\includegraphics[width=\linewidth,height=7.0cm]{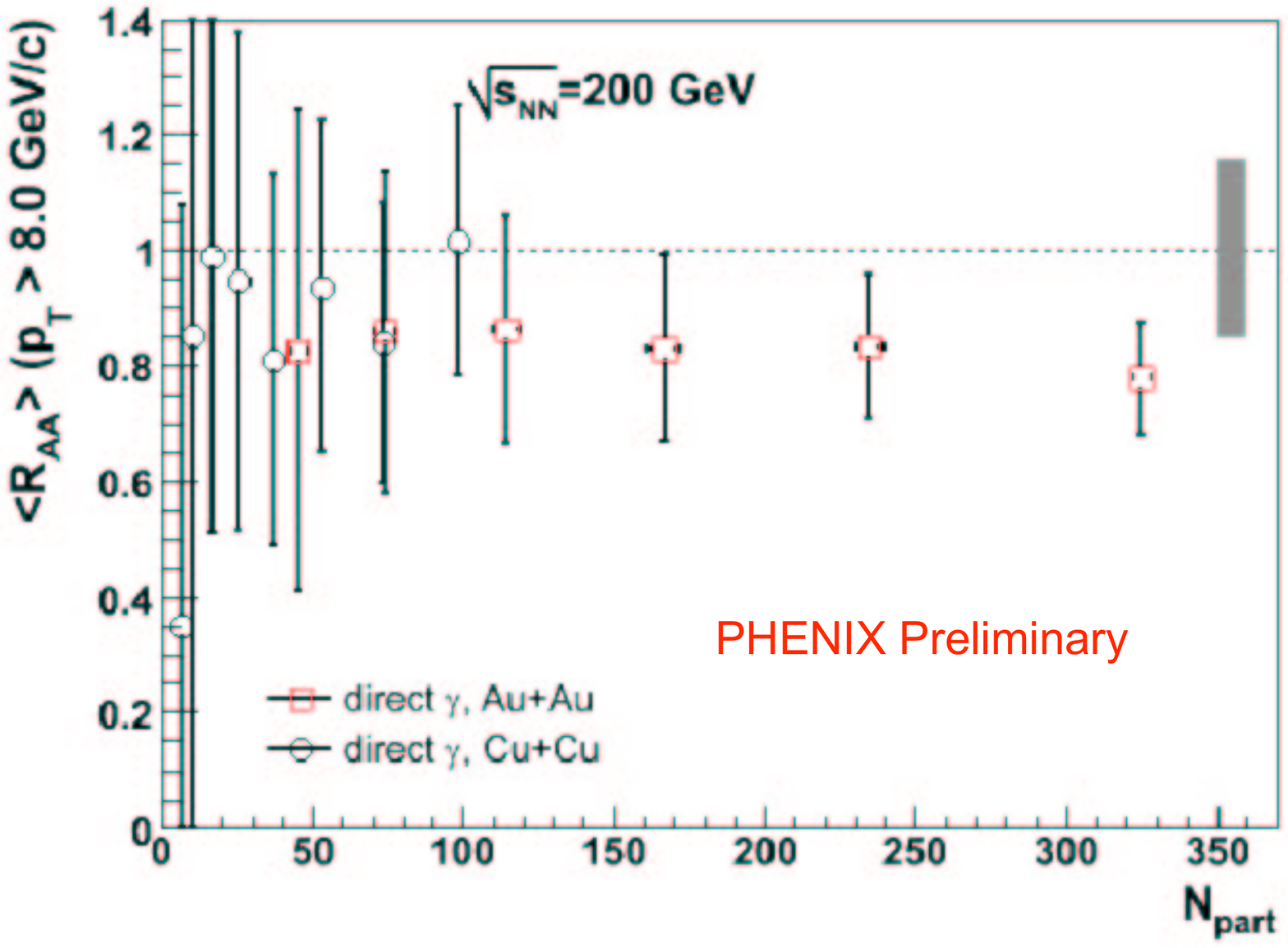}
		\caption{{Nuclear modification factor R$_{AA}$ for high $p_T$ direct $\gamma$ production as a function of the number of participant nucleons in Cu+Cu and Au+Au collisions. }} 
		\label{fig:CuAuGamRAA}
	\end{minipage}
\end{figure}

New PHENIX results from Run 5 indicate that the neutral pion suppression is similar for Cu+Cu and Au+Au collisions for centrality selections with the same number of participating nucleons, as shown in Figure~\ref{fig:CuAuRAA}~\cite{ppg084}. On closer inspection, systematic differences between the Cu+Cu and Au+Au systems do appear at small participant numbers where differences in the  geometry of the nuclear overlap may become important. 

The dependence of the pion suppression on collision energy ($\sqrt{s_{NN}}$) has also been investigated with the Cu+Cu $\pi^0$ measurements where it is seen in Figure~\ref{fig:CuRAA} that the suppression is quite similar at 200 and 62.4 GeV, but that the $\pi^0$ yield is instead enhanced compared to expectations from p+p collisions at 22.4 GeV, which  indicates that suppression due to parton energy loss begins to dominate over Cronin enhancement between 22 and 62 GeV~\cite{ppg084}.  The $\sqrt{s_{NN}}$ dependence of the observed suppression is in good agreement with parton energy loss calculations~\cite{Vitev}.

As expected from previous observations,  the high $p_T$ direct photon yields are consistent within errors with no suppression for all centralities for both Cu+Cu and Au+Au collitions, as shown in Figure~\ref{fig:CuAuGamRAA}. The systematic study of the measured high $p_T$ suppression of the single particle yields with varying particle species type (including heavy flavor), collision system centrality and $\sqrt{s_{NN}}$ will provide important input to model descriptions from which information about the opacity of the produced matter may be deduced by quantitative comparisons of the model predictions with the data~\cite{ppg079,Reygers}. Beyond inclusive particle measurements, the parton energy loss is being further investigated through studies of high $p_T$ direct photon $v_2$ measurements~\cite{Miki} and $\pi^0$ production as a function of the orientation with respect to the reaction plane~\cite{PHv2RAA}, via two- and three-hadron correlations~\cite{Franz} as well as through $\gamma$-hadron correlations~\cite{Ngyuen}.

\section{J/$\psi$ Production} \label{sec:jpsi}

\begin{figure}[h]
	\begin{minipage}[t]{0.5\linewidth}
		\centering
		\includegraphics[width=\linewidth,height=8.0cm]{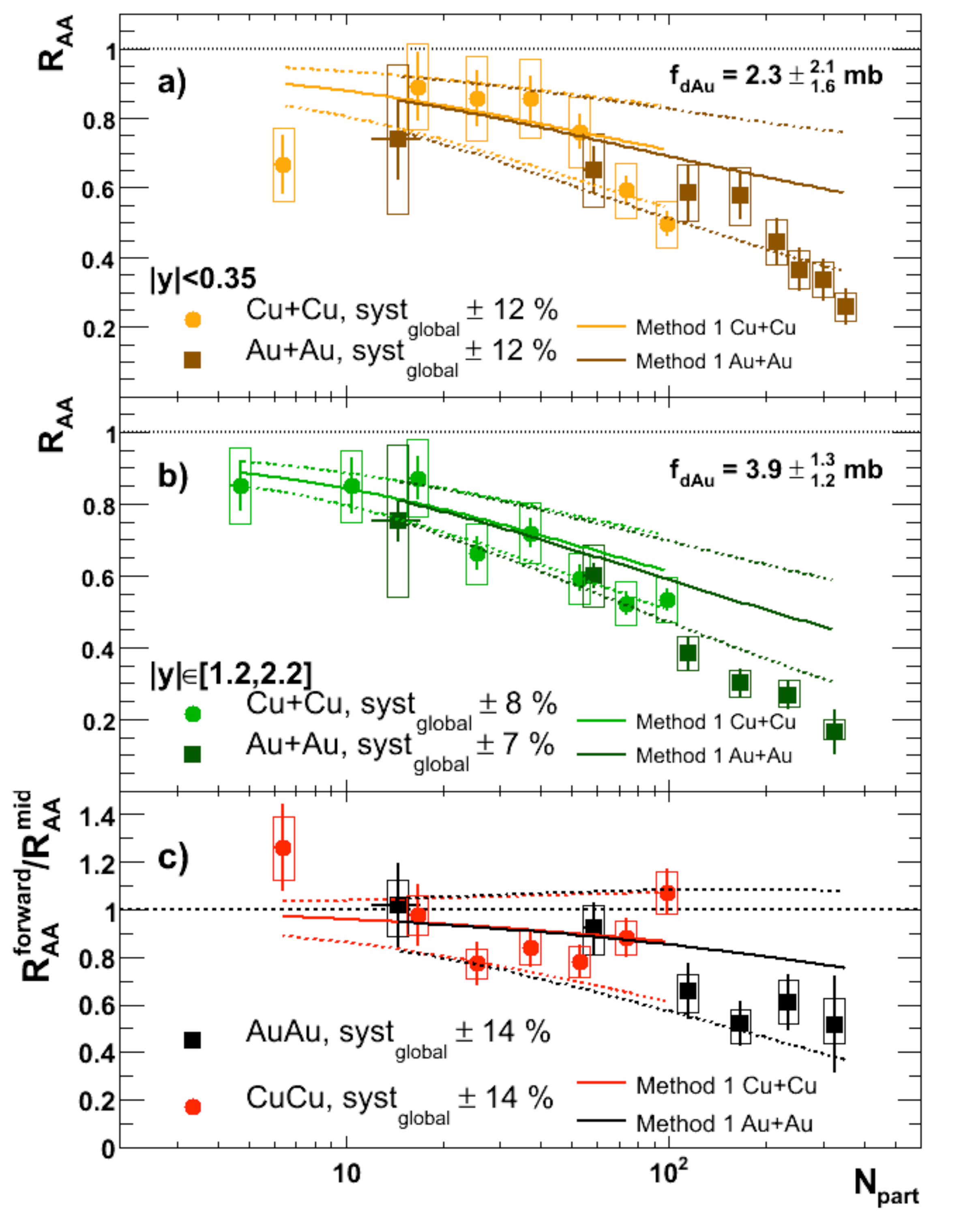}
		\caption{{Nuclear modification factor R$_{AA}$ as a function of the number of participant nucleons for Au+Au and Cu+Cu collisions at a) mid-rapidity and b) forward rapidity, and c) ratio. }} 
		\label{fig:CuAuJpsiRAA}
	\end{minipage}
	\begin{minipage}[t]{0.5\linewidth}
		\centering
		\includegraphics[width=\linewidth,height=7.0cm]{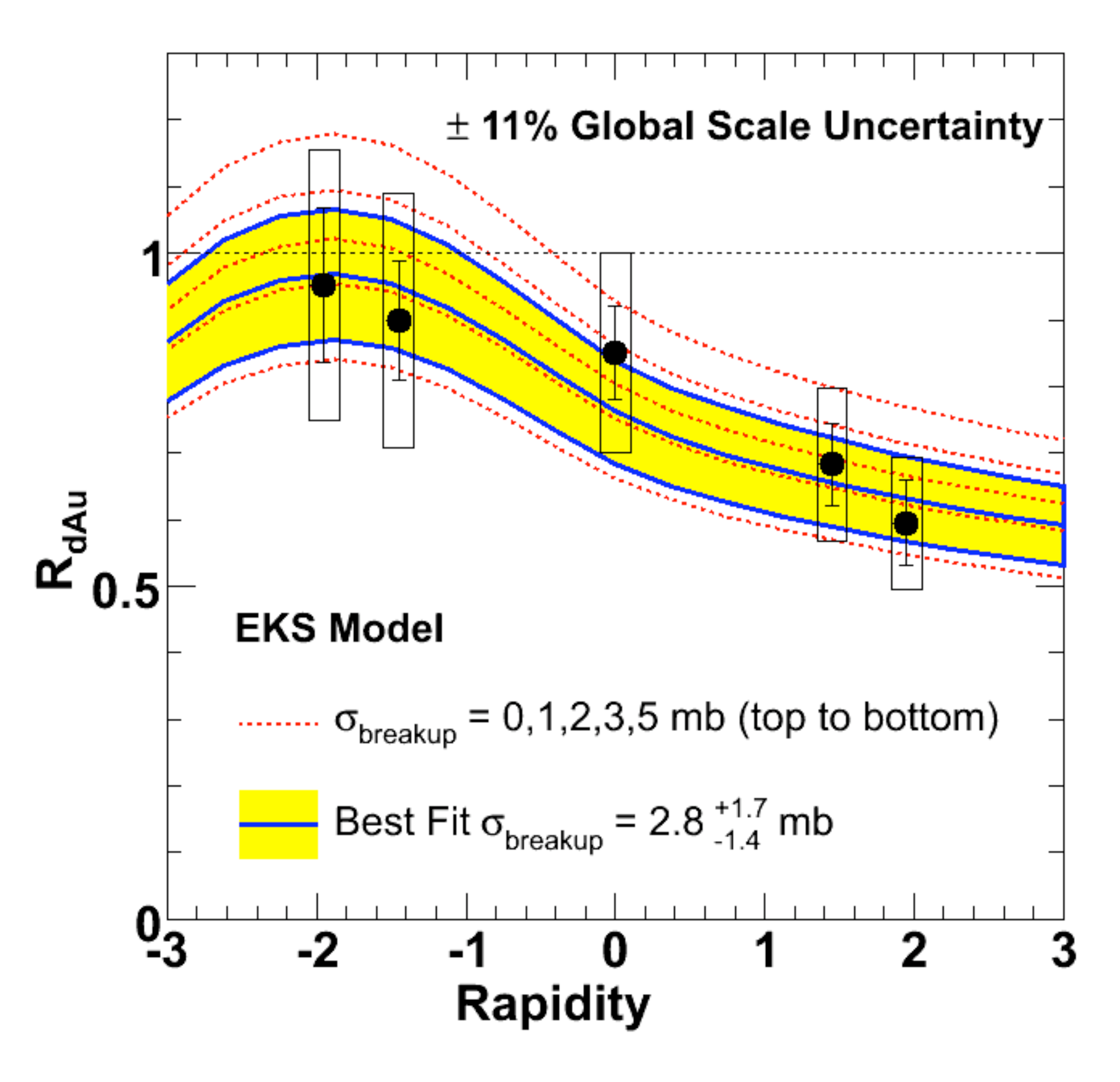}
		\caption{{Nuclear modification factor R$_{dAu}$ as a function of rapidity for $J/\psi$ production in d+Au collisions compared to EKS model calculations~\cite{EKS}.}}
		\label{fig:dAuJpsiRAA}
	\end{minipage} 
\end{figure}

One of the earliest proposed signatures of the formation of dense deconfined matter (Quark Gluon Plasma) in relativistic heavy ion collisions was the predicted suppression of  the J/$\psi$ yield due to Debye screening of the $c\overline{c}$ quark bound state in the dense partonic matter~\cite{JPsisup}.  Suppression of the $J/\psi$ yield was soon afterwards observed in measurements at the CERN SPS~\cite{NA50sup}.  New results from PHENIX~\cite{Oda} shown in Figure~\ref{fig:CuAuJpsiRAA} indicate that the J/$\psi$ suppression is the same for Cu+Cu and Au+Au collisions for centrality selections with the same number of participating nucleons~\cite{ppg071}. The amount of suppression increases with the number of participating nucleons and surprisingly is rather similar at RHIC and SPS energies~\cite{PHjpsi}.  Also, contrary to expectations from Debye screening alone, the suppression is observed to be stronger at forward rapidity than at mid-rapidity (Figure~\ref{fig:CuAuJpsiRAA}).

At SPS energies the J/$\psi$ yield is suppressed also in p+A collisions. This is interpreted as a Òcold nuclear matterÓ (CNM) effect as a result of modification of the nucleon parton momentum distributions in the nucleus and the breakup of the J/$\psi$ due to its interaction with the cold spectator nucleons.  PHENIX has investigated the cold nuclear matter effects on J/$\psi$ suppression at RHIC energies using d+Au collisions, with comparison to p+p collisions.
A new analysis was recently completed using the Run 5 p+p data set, which is a factor of 10 larger than the Run 3 p+p data set, together with the Run 3 d+Au data set. An improved analysis with better understanding of detector effects has been applied consistently to both data sets. As shown in Figure~\ref{fig:dAuJpsiRAA} the new result on the d+Au nuclear modification factor $R_{dAu}$ still has a rather large uncertainty that prevents to draw firm quantitative statements on any additional suppression in Au+Au collisions beyond cold nuclear matter effects (see Figure~\ref{fig:CuAuJpsiRAA})~\cite{PHdAuJpsi,Wysocki}. With the PHENIX Run 8 d+Au data sample just obtained it is estimated that the number of J/$\psi$'s accumulated is a factor of 50 greater than for Run 3, which should allow to constrain the contribution from cold nuclear matter effects more strongly.

\section{Heavy Flavor} \label{sec:heavy}

\begin{figure}[h]
	\begin{minipage}[t]{0.5\linewidth}
		\centering
		\includegraphics[width=\linewidth]{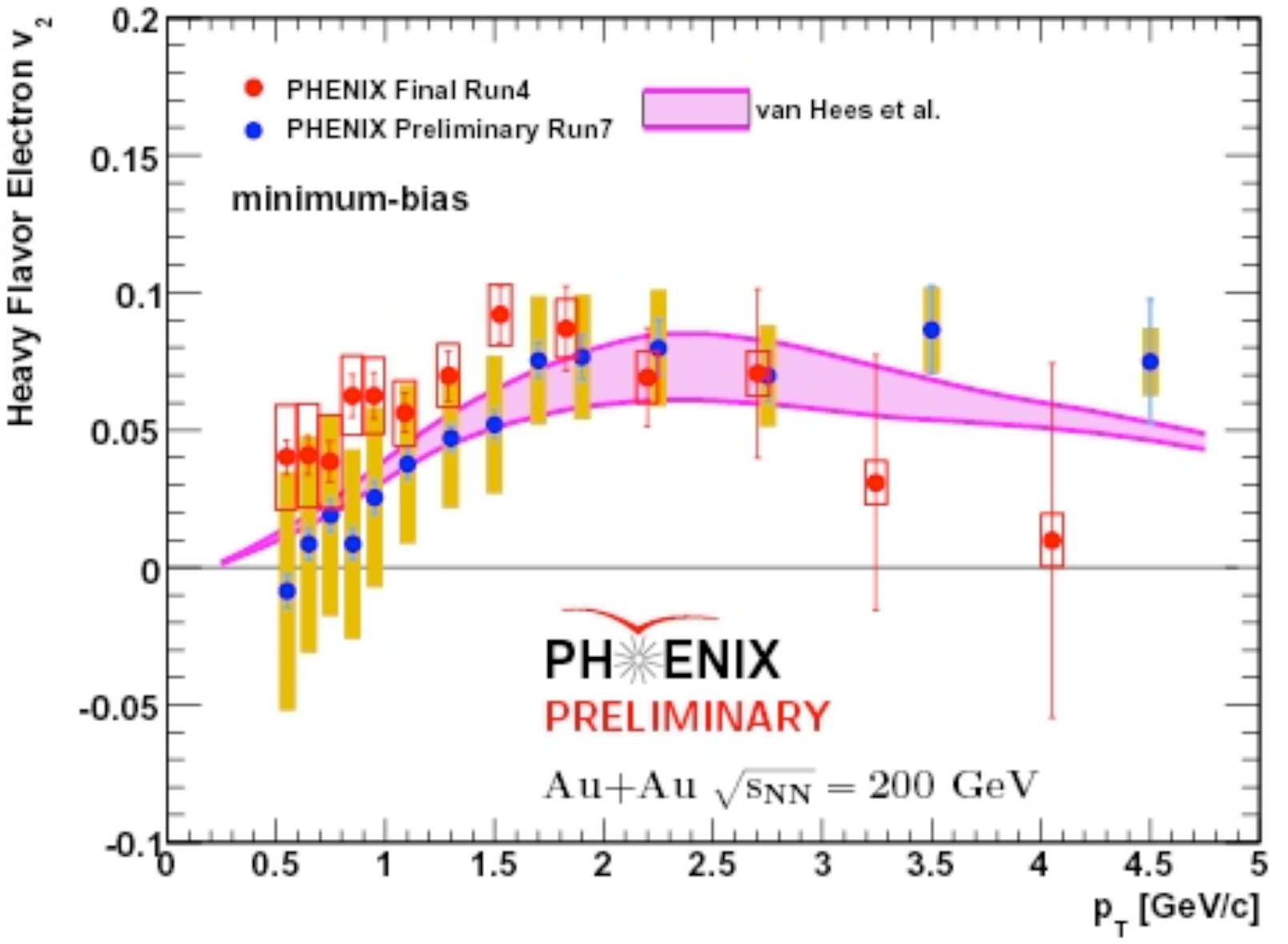}
		\caption{{ Azimuthal anisotropy parameter $v_2$ of heavy flavor electrons in minimum bias Au+Au collisions from Run 4 ~\cite{PHhf} and  new preliminary results from Run 7 compared to transport calculations by van Hees~\cite{vHees}.}}
		\label{fig:HeavyFlow}
	\end{minipage}
	\begin{minipage}[t]{0.5\linewidth}
		\centering
		\includegraphics[width=\linewidth]{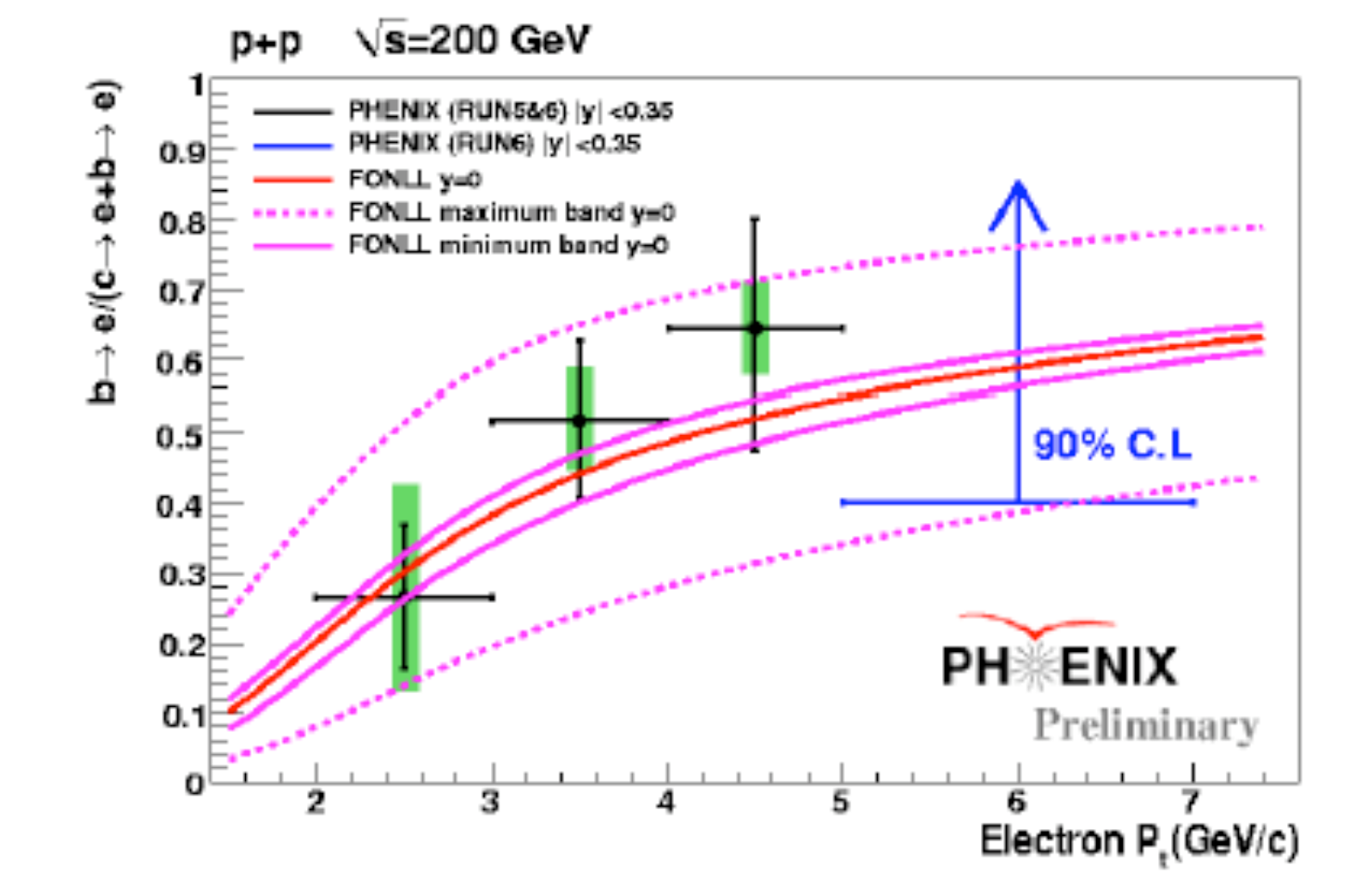}
		\caption{{ Ratio of electrons from bottom to electrons from charm compared to FONLL calculations~\cite{FONLL}.}}
		\label{fig:bcRatio}
	\end{minipage}
\end{figure}

First measurements on heavy flavor quark production in PHENIX via single electron measurements provided another surprising result at RHIC~\cite{PHhf}.  Although it was expected that heavy quarks should show less stopping than light quarks in dense partonic matter, and hence heavy quark jets should show less quenching,  it was instead observed that the yield of electrons from heavy flavor at high transverse momentum was suppressed in Au+Au collisions with nearly the same suppression as observed for $\pi^0$'s. In addition, single electrons attributed to heavy flavor were observed to show large azimuthal asymmetries~\cite{PHhf},  with the single electron $v_2$ results recently extended  to higher transverse momenta with the Run 7 Au+Au data set, as shown in Figure~\ref{fig:HeavyFlow}. Taken together, the results implied significant damping of the motion of heavy quarks as they propagate through the dense matter produced at RHIC. Together with model calculations these observations allow to extract information on the viscosity to entropy ratio, $\eta/s$, of the dense matter and draw the conclusion that the matter has an $\eta/s$ ratio close to the conjectured lower bound of $\sim 1/4\pi$, essentially a "perfect liquid"~\cite{perfectliquid,Averbeck}.

With the goal to better separate the charm and bottom contributions (and shed light on the unresolved discrepancies between the STAR and PHENIX charm results) PHENIX has performed single muon measurements at forward rapidities in p+p collisions to extract information on the rapidity dependence of the heavy flavor production~\cite{Hornback}. Also, electron-hadron correlation measurements have been used to perform a pseudo-$D^0$ analysis to explicitly separate the charm and bottom contributions~\cite{Morino} in p+p collisions with results shown in Figure~\ref{fig:bcRatio}. The measured ratio is found to be in good agreement with FONLL calculations, although separately the charm and bottom measurement each disagree significantly with the FONLL calculations~\cite{Morino}.

\section{Low Mass Electron Pairs} \label{sec:electrons}

\begin{figure}[h]
	\begin{minipage}[t]{0.5\linewidth}
		\centering
		\includegraphics[width=\linewidth,height=7.0cm]{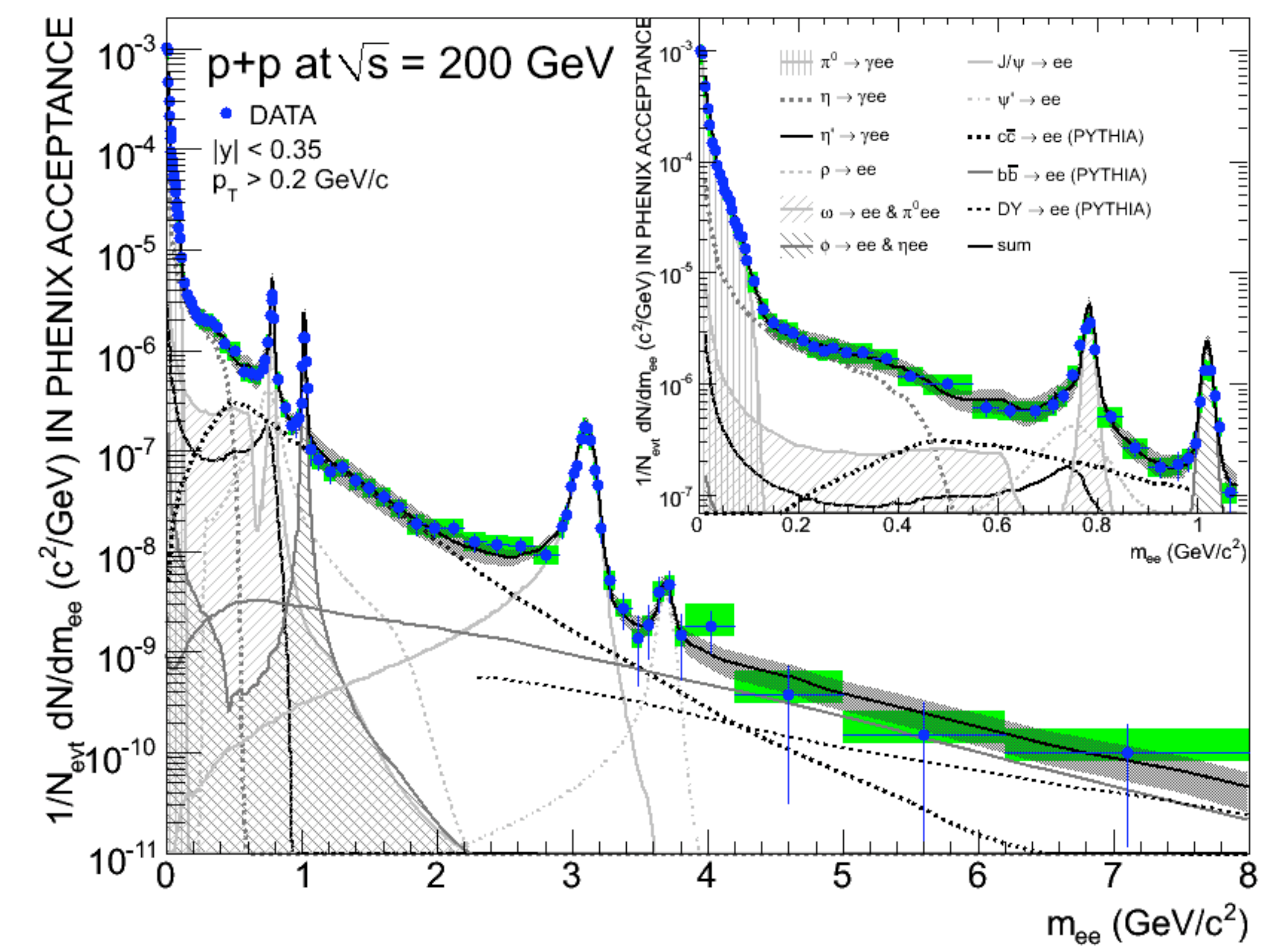}
		\caption{{The yield of $e^+e^-$ pairs per p+p collision compared to the yield expected from hadronic decays. Statistical (bars) and systematic (boxes) uncertainties are shown separately.}}
		\label{fig:ppMee}
	\end{minipage} 
	\begin{minipage}[t]{0.5\linewidth}
		\centering
		\includegraphics[width=\linewidth,height=7.0cm]{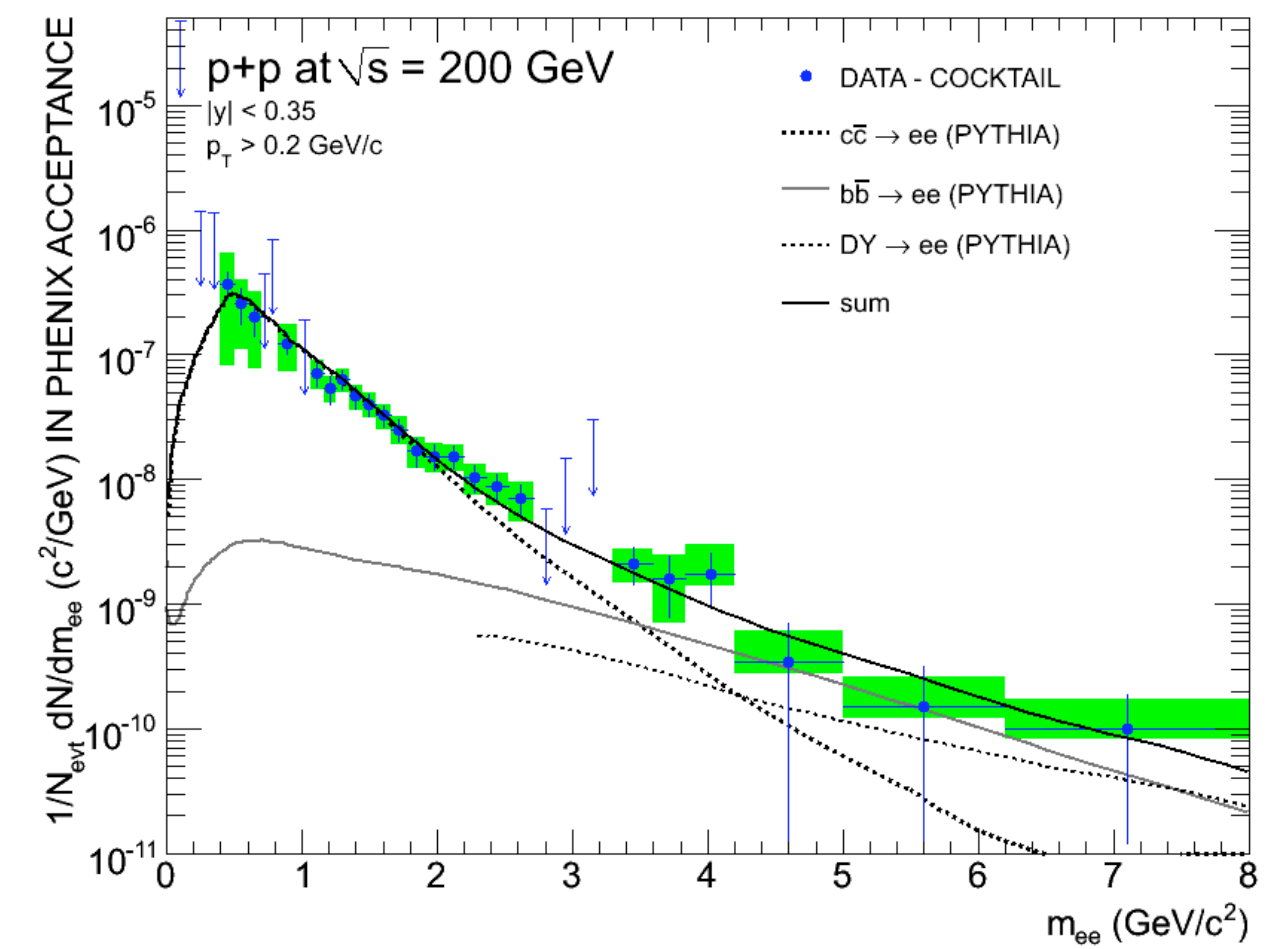}
		\caption{{The yield of $e^+e^-$ pairs after subtraction of the contribution of hadronic decays. }} 
		\label{fig:ppCandB}
	\end{minipage}
\end{figure}

The measurement of electron-positron pairs allows the study of direct virtual photon production with the advantage that the main background from neutral pion decays (due to internal or external photon conversion) is about two orders of magnitude smaller than for the measurement of real direct photons. An additional order of magnitude reduction of the background to the virtual photon measurement can be obtained by using pairs with mass above the pion mass.  The $e^+e^-$ pair mass spectrum for p+p collisions at 200 GeV from the PHENIX Run 5 data set~\cite{ppg085} is  shown in Figure~\ref{fig:ppMee}. It is seen to be  in very good agreement with the expected yield based on a Monte Carlo calculation of the electron-pair yield from hadron decays using measured hadron yields, together with contributions to the $e^+e^-$ mass spectrum from Drell-Yan, $c\overline{c}$, and $b\overline{b}$ decays as predicted by the PYTHIA model. Figure~\ref{fig:ppCandB} shows the measured $e^+e^-$ mass spectrum, after subtraction of the hadronic decay contribution, compared separately to the PYTHIA model predictions to demonstrate the good agreement of the measurement with the predicted $c\overline{c}$ and $b\overline{b}$ contributions.

\begin{figure}[h]
	\begin{minipage}[t]{0.5\linewidth}
		\centering
		\includegraphics[width=\linewidth,height=7.0cm]{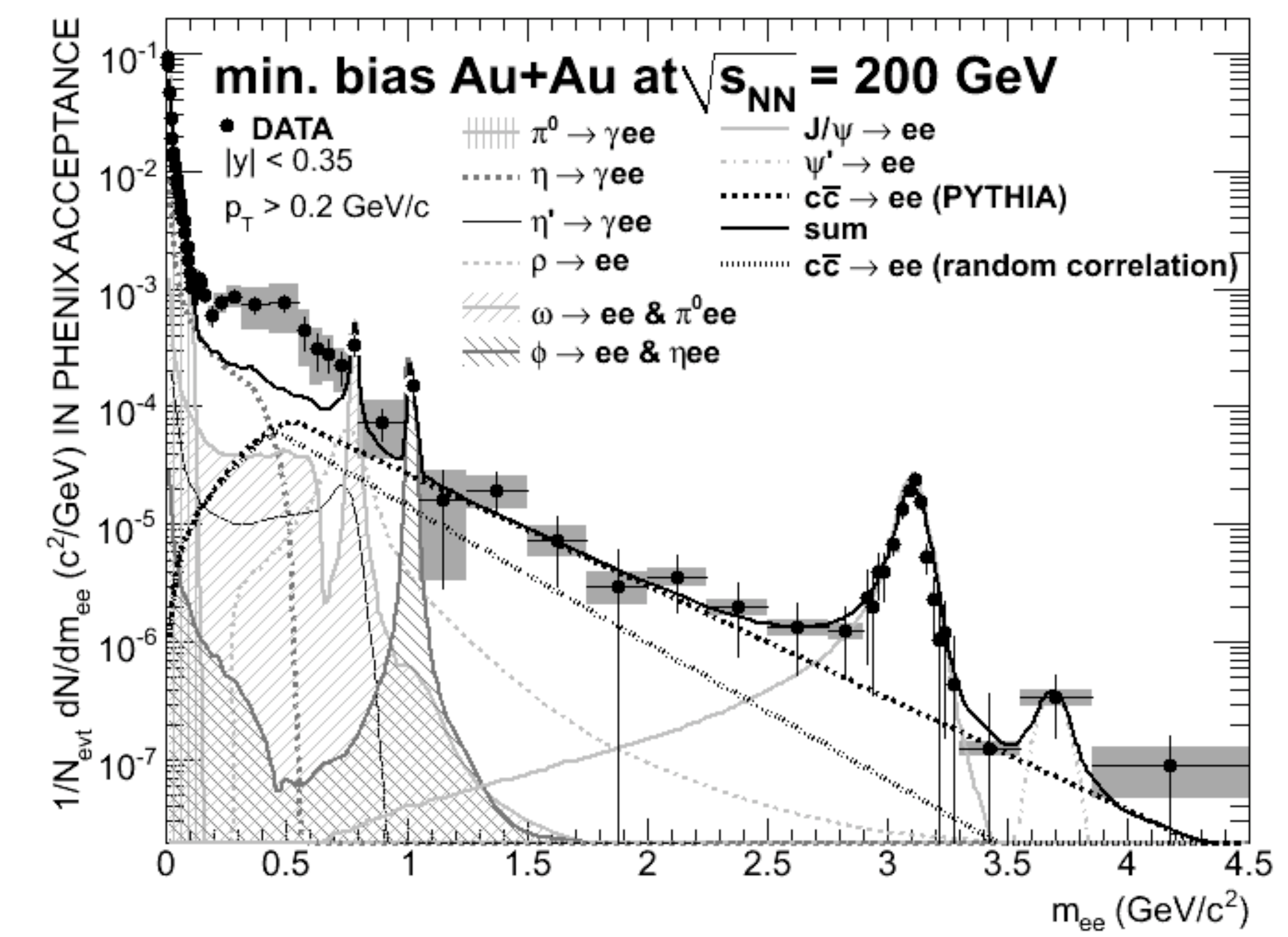}
		\caption{{Invariant yield of $e^+e^-$ pairs in minimum bias Au+Au collisions compared to the yield expected from hadronic decays. Statistical (bars) and systematic (boxes) uncertainties are shown separately.}}
		\label{fig:AuAuMee}
	\end{minipage} 
	\begin{minipage}[t]{0.5\linewidth}
		\centering
		\includegraphics[width=\linewidth,height=7.5cm]{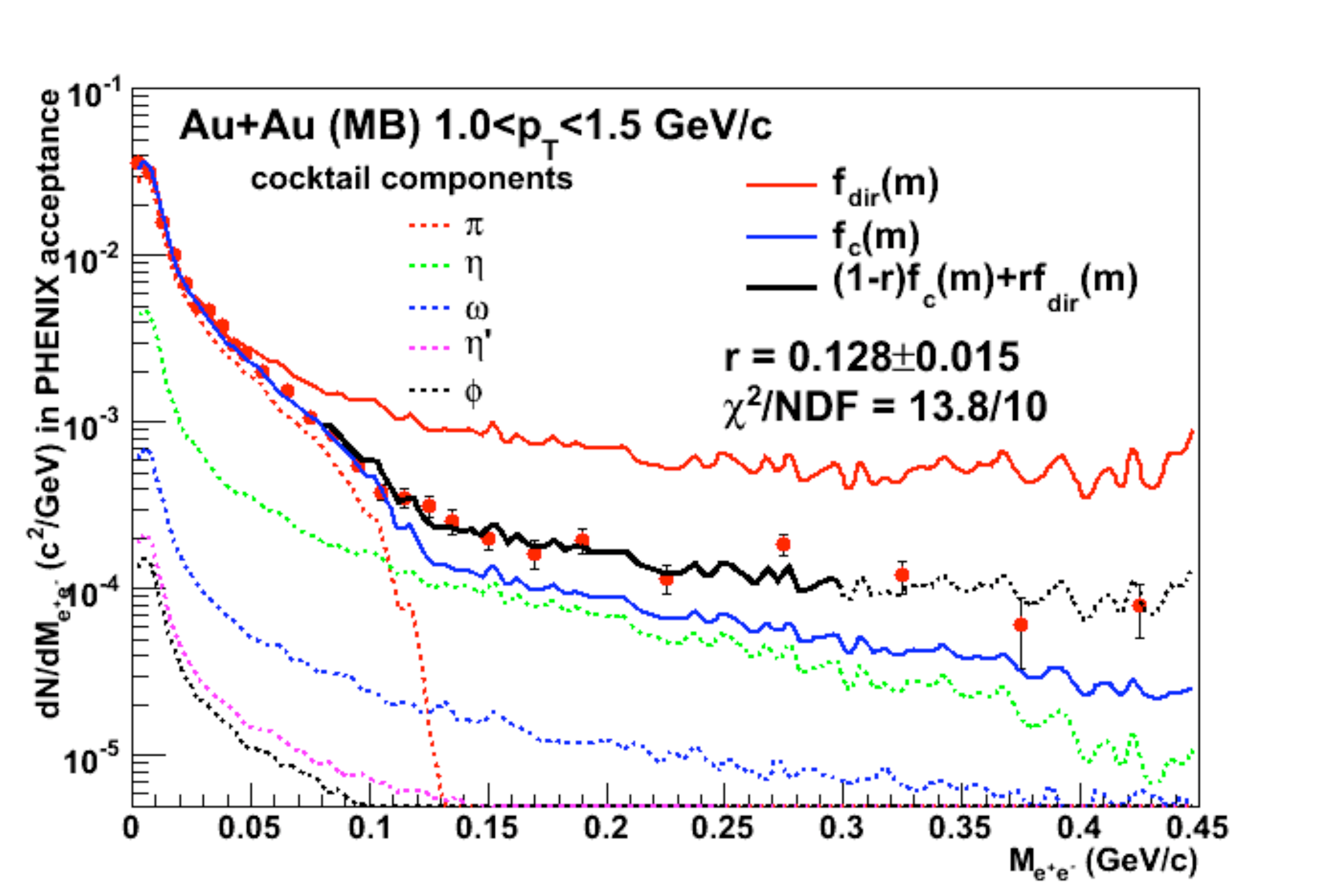}
		\caption{{Invariant yield of low mass $e^+e^-$ pairs with $1.0 < p_T < 1.5$ GeV/c in minimum bias Au+Au collisions. The solid blue curve indicates the expectation with the measured $\eta$ Dalitz contribution. }} 
		\label{fig:AuAuExcess}
	\end{minipage}
\end{figure}

In the case of Au+Au collisions, the $e^+e^-$ mass spectrum shows a very large excess beyond expectations from hadronic decays in the low mass region between the $\pi^0$ and $\omega$ meson masses, as shown in Figure~\ref{fig:AuAuMee}~\cite{ppg075}. This excess is dominantly at transverse momenta below about 1 GeV/c, indicating that it is produced in the cooler late hadronic phase of the collision. 

\begin{figure}[h]
	\begin{minipage}[b]{\linewidth}
		\centering
		\includegraphics[width=0.6\linewidth]{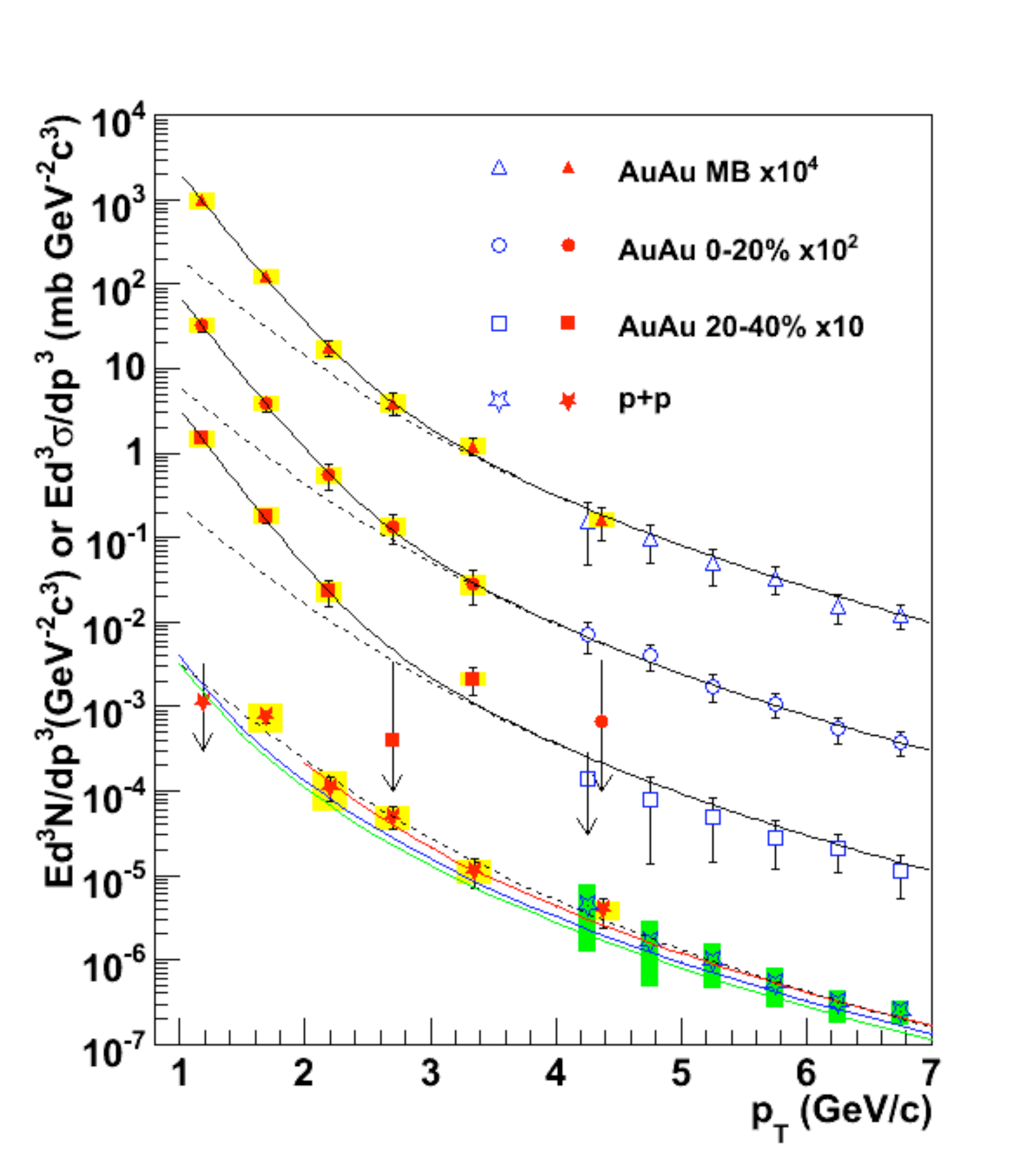}
		\caption{{Invariant cross section of direct photons in p+p collisions and Au+Au collisions for 	
		several centrality selections compared to scaled NLO pQCD 
		predictions (dashed curves). The open points are previously published PHENIX 
		results~\cite{PHgam,PHppgam}. }}
		\label{fig:GamExcess}
	\end{minipage}
\end{figure}

However, a significant $e^+e^-$ excess also persists  at higher transverse momenta in the $e^+e^-$ mass region above the  $\pi^0$ mass, as seen in Figure~\ref{fig:AuAuExcess}. This excess can be used to extract the virtual photon momentum spectrum with an error significantly smaller, in the low transverse momentum region, than obtained by measurement of real photons~\cite{PHgam,PHppgam,ppg086,Dahms}. A small but significant excess is also observed in p+p collisions in the mass region above the $\pi^0$ mass at transverse momenta above 1 GeV/c. As shown in Figure~\ref{fig:GamExcess}, the measured invariant photon yield for p+p collisions by the virtual photon measurement is found to be consistent with expectations from pQCD predictions.

On the other hand, for Au+Au collisions, the virtual photon yield associated with the observed $e^+e^-$ excess is greater than that expected from the p+p measurement, which  suggests that it is due to thermal radiation from the early phase of the Au+Au collision. These measurements hold promise that the thermal photon spectrum may finally be extracted with sufficient precision to provide significant constraints on the initial temperature of the dense matter being created at RHIC~\cite{ppg086,Dahms,Stankus}.


\end{document}